\pdfoutput=1
\documentclass{article}
\usepackage{graphicx, indentfirst, hyperref}
\usepackage[a4paper, margin = 3cm]{geometry}
\usepackage[numbers]{natbib}
\makeatletter
\let\mailmark\@fnsymbol
\makeatother

\newcommand*{\cf}{\emph{cf.}}
\newcommand*{\eg}{\emph{eg.}}
\newcommand*{\etc}{\emph{etc}}
\newcommand*{\prog}[1]{\emph{#1}}
\newcommand*{\figref}[1]{Figure \ref{fig:#1}}
\newcommand*{\secref}[1]{Section \ref{sec:#1}}
\let\thxmark\textsuperscript
\begin{document}
\title{%
	PandA(Box) flies on \prog{Bluesky}: maintainable and
	user-friendly fly scans with \prog{Mamba} at HEPS%
}
\author{%
	Peng-Cheng Li\thxmark{1}, Cheng-Long Zhang\thxmark{1},
	Yu-Jun Zhang\thxmark{1}, Chun Li\thxmark{1}, Zhi-Ying Guo\thxmark{1},\\
	Ge Lei\thxmark{1,2}, Yi Zhang\thxmark{1,2}, Ai-Yu Zhou\thxmark{1},
	Xiao-Xue Bi\thxmark{1,\mailmark{1}}, Yu Liu\thxmark{1,\mailmark{1}}%
}
\date{}
\maketitle
\begingroup
\renewcommand{\thefootnote}{\fnsymbol{footnote}}
\footnotetext[1]{\ %
	Correspondence e-mail:
	\texttt{bixx@ihep.ac.cn}, \texttt{liuyu91@ihep.ac.cn}.%
}
\endgroup
\footnotetext[1]{\ %
	Institute of High Energy Physics, Chinese Academy of Sciences,
	Beijing 100049, People's Republic of China.%
}
\footnotetext[2]{\ %
	University of Chinese Academy of Sciences,
	Beijing 100049, People's Republic of China.%
}

\section*{Abstract}

\paragraph{Keywords:}
Fly scan; PandABox; \prog{Bluesky}; \prog{Mamba}; XRF mapping.

\paragraph{Purpose:}
Fly scans are indispensible in many experiments at the
High Energy Photon Source (HEPS).  PandABox, the main platform
to implement fly scans at HEPS, needs to be integrated into \prog{Mamba},
the experiment control system developed at HEPS based on \prog{Bluesky}.

\paragraph{Methods:}
In less than 600 lines of easily customisable and extensible backend code,
provided are full control of PandABox's TCP server in native \prog{ophyd},
automated configuration (also including wiring) of ``PandA blocks'' for
constant-speed mapping experiments of various dimensions, as well as
generation of scans deliberately fragmented to deal with hardware
limits in numbers of exposure frames or sequencer table entries.

\paragraph{Results:}
The upper-level control system for PandABox has been ported to
\prog{Bluesky}, enabling the combination of both components' flexibility
in fly-scan applications. Based on this backend, a user-friendly
\prog{Mamba} frontend is developed for X-ray fluorescence (XRF)
mapping experiments, which provides fully online visual feedback.

\section{Introduction}\label{sec:intro}

At the High Energy Photon Source (HEPS) \cite{jiao2018}, a 4th-generation
synchrotron radiation facility, because of its small light spots and high
brightness, fly scans (continuous scans) are a mandatory requirement to reduce
the scanning time and radiation damage in beamline experiments.  Among the
off-the-shelf hardware controllers available, the PandABox \cite{zhang2017}
has been chosen as the main platform to implement fly scans at HEPS.  We find
this choice appropriate because of PandABox's desirable hardware attributes,
which make it usable out of the box or with minor modifications in perhaps
$>90\%$ applications at HEPS (at least its Phase I): FPGA logics encapsulated
in ``P(osition) and A(cquisition)'' blocks (\cite{dls2015a}; \cf\ also
\figref{blocks}) understandable to programmers without a strong background
in hardware engineering, block configuration (including wiring) stored
in registers (instead of FPGA interconnections) enabling fast
reconfiguration, as well as open-source PandA blocks and
open hardware designs allowing for deeper customisation.

\begin{figure}[htbp]\centering
\includegraphics[width = 0.9\textwidth]{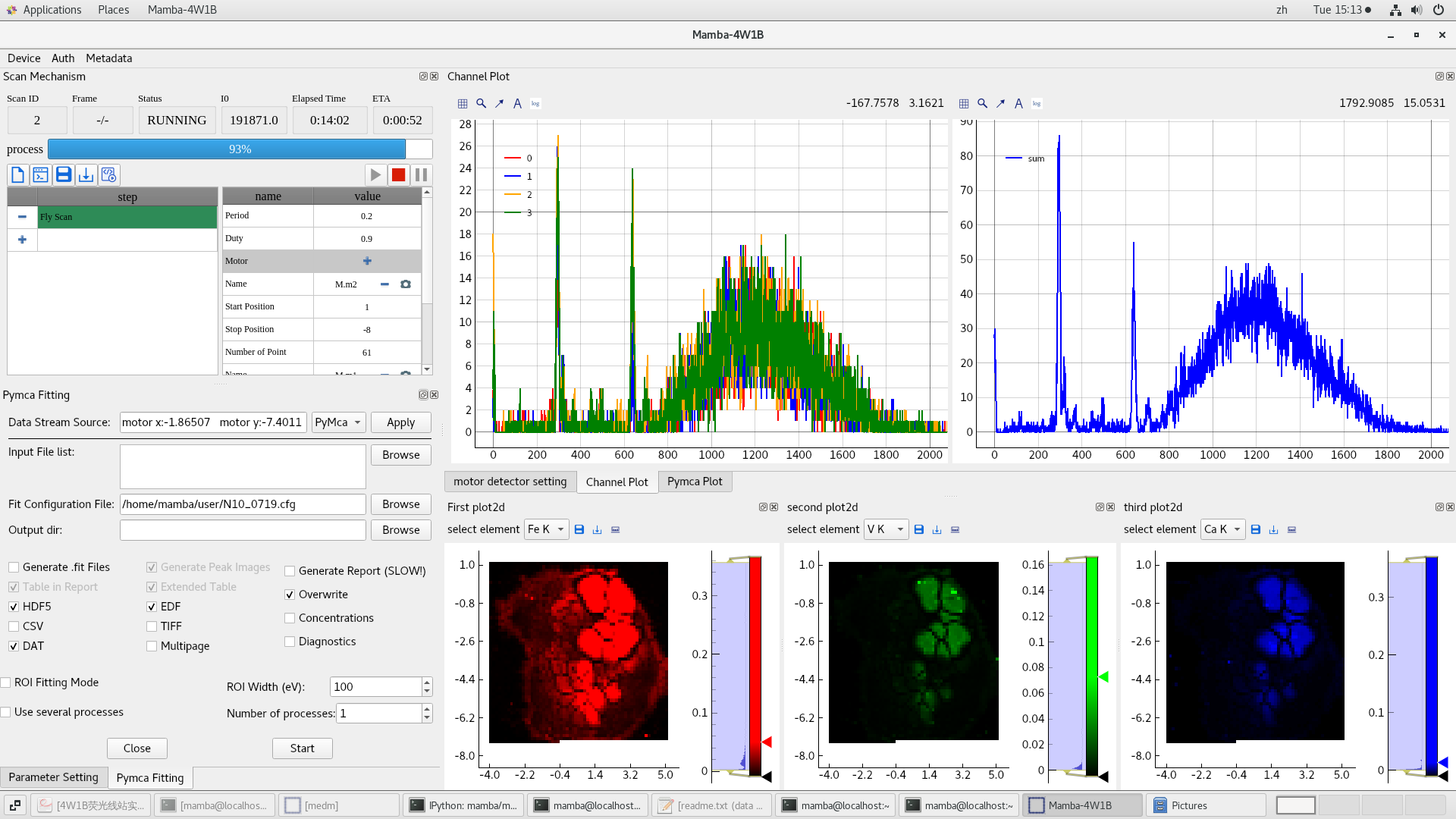}
\caption{\prog{Mamba} frontend for XRF mapping}\label{fig:gui}
\end{figure}

However, despite the hardware versatility of PandABox, after early assessment
of its official middleware, \prog{pymalcolm} \cite{basham2018}, we find several
issues with the latter: vastly different architecture from \prog{Bluesky}
\cite{allan2019} and consequently \prog{Mamba} \cite{liu2022, dong2022}, which
are the fundamental software frameworks for beamline experiments at HEPS;
quite inexpressive, inflexible and indirect configuration, conflicting with
the notion of experiment parameter generators (EPGs, \cf\ \secref{choice})
in \prog{Mamba}; lack of consideration about online data processing, necessary
for user-friendly status feedback and timely error detection.  Instead, by
only reusing \verb|pandablocksclient.py| from \prog{pymalcolm}, we are able
to directly encapsulate PandABox's interface in \prog{Bluesky}'s hardware
abstraction layer, \prog{ophyd} \cite{bnl2014}, and implement a backend EPG
for constant-speed scans of various dimensions.  The backend is, excluding
\verb|pandablocksclient.py|, less than 600 lines, and easily extensible to
more complex fly scans; less code implies less development/maintenance cost,
and the majority of its code was written by one developer in 3 weeks.  Based
on the backend, we have developed a user-friendly \prog{Mamba} frontend
for X-ray fluorescence (XRF) mapping (\figref{gui}), with fully online
feedback providing both spectra generated by \prog{PyMca} \cite{sole2007}
and the map in progress.  The frontend has been deployed at the 4W1B
beamline of Beijing Synchrotron Radiation Facility (BSRF), and has
seen satisfactory adoption from users -- our fly-scan mode
has almost replaced the old step-scan mode.

\section{Some notes on the choice of technologies}\label{sec:choice}

As has been mentioned in \secref{intro}, we have made the unorthodox decision
to integrate PandABox into \prog{Bluesky} without using \prog{pymalcolm}
(other than \verb|pandablocksclient.py|).  Actually, we have also avoided
\prog{ophyd2} \cite{forrester2022} co-developed by the authors of \prog{Bluesky}
and \prog{pymalcolm}; moreover, as will be shown later, we have deviated from
\prog{Bluesky}'s officially suggested \verb|kickoff|/\verb|complete|/%
\verb|collect| paradigm for fly scans.  These decisions deserve
some explanation, which is the subject of this section.

Essentially, we hope our requirements can be implemented in succinct code, so
we want libraries to be \emph{expressive}; we hope the libraries can be easily
adapted to reasonable yet unanticipated requirements, so we want libraries to
be \emph{flexible}; we hope the libraries can be easily debugged and extended,
so we want libraries to be \emph{straightforward} (\emph{direct}).  These are
exactly the fundamental ideas behind \prog{Mamba}'s notion of EPGs, which were
originally meant to remove unnecessary duplication in user inputs for beamline
experiments.  Now they have been expanded to simplify programming on multiple
levels: users, beamline operators, and even developers themselves, in order
to reduce maintenance cost (often severalfold or even more than tenfold
simplification) while keeping (if not enhancing) user friendliness.  Users
unfamiliar with programming can also benifit from EPGs, as \emph{Mamba}'s
command-injection mechanism allows frontends to fully exploit potentials
of the backend command line, including the expressiveness of EPGs.

\prog{pymalcolm} uses YAML configuration files, which are fairly verbose
and not dynamic enough for programmatic manipulation; there seems to be no
further abstraction for configurations of PandA blocks or \prog{pymalcolm}'s
own blocks/methods/attributes, either.  We attempted to envision features
like scan fragmentation, ETA estimation \etc\ in \prog{pymalcolm}'s way, but
deemed it too complex for us.  We also find the \prog{pymalcolm} internals
non-trivial to understand, but that may be because of some stylistic
difference from \prog{Bluesky}.  By contrast, our fly-scan backend EPG allows
experiment-specific configuration to be often just a couple dozens of lines
of code; it supports advanced features we want and is quite extensible;
it is less than 600 lines of quite well-organised code.  Even in comparison
with \prog{ophyd2} and taking into account the cost of self-maintenance,
we find our backend EPG to offer more benefits at less costs.

Speaking of \prog{Bluesky}'s \verb|kickoff|/\verb|complete|/\verb|collect|
paradigm, our main problem with it is the lack of facility for online data
processing.  Surely we could set up a thread that handles incoming data in the
background, but the \verb|complete|/\verb|collect| parts would still become
nearly meaningless (apart from acting as some kind of stop notification).
Instead, we devised our solution based on direct access to data files through
the SWMR extension of HDF5; this, along with the advanced features mentioned
above, is discussed in \secref{bextra}.

\section{Encapsulating PandABox's interface with \prog{ophyd}}

\begin{figure}[htbp]\centering
\includegraphics[width = 0.5\textwidth]{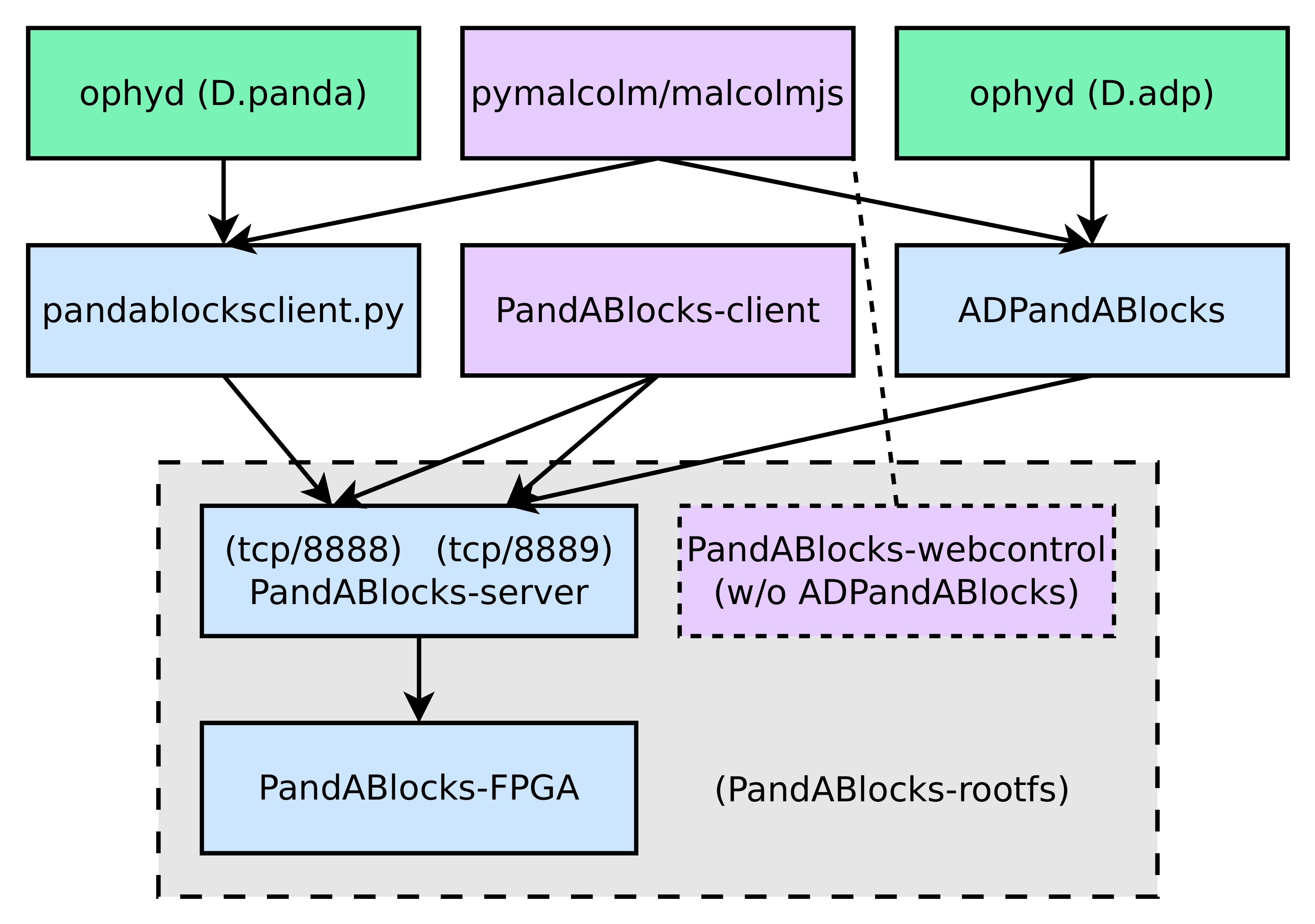}
\caption{%
	PandABox's software communication.  \prog{PandABlocks-client} is an
	official client capable of both control and data access, written in Python
	but independent of \prog{pymalcolm}; we do not use it as of now, simply
	because \texttt{pandablocksclient.py} is shorter and we can already access
	\prog{ADPandABlocks} from \prog{ophyd}.  \prog{PandABlocks-webcontrol}
	is basically a trimmed-down \prog{pymalcolm} (but still including
	the \prog{malcolmjs} Web interface), which we currently do not
	plan to replace as it works well enough for us out of the box.
	The \texttt{M.$\langle$\prog{motor}$\rangle$}/%
	\texttt{D.$\langle$\prog{detector}$\rangle$}
	convention was introduced by \prog{Mamba}.%
}\label{fig:fwsw}
\end{figure}

\begin{figure}[htbp]\centering
\includegraphics[width = 0.5\textwidth]{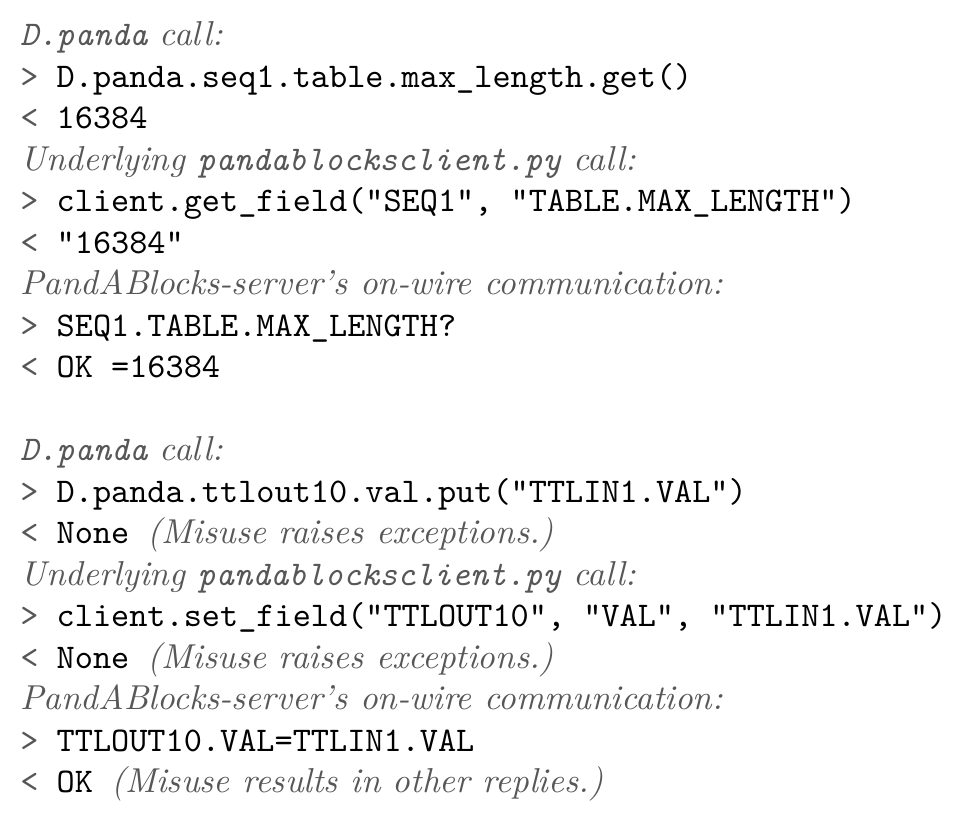}
\caption{%
	Usage comparison between our \prog{ophyd} module for PandABox,
	\texttt{pandablocksclient.py} and \prog{PandABlocks-server}%
}\label{fig:pbccomp}
\end{figure}

In order to satisfactorily reimplement \prog{pymalcolm}'s fly-scan
functionalities in \prog{Bluesky}, it is highly instructive to analyse the
communication between components (and especially those around PandABox) in
\prog{pymalcolm}-based fly scans.  As the hardware communication is fairly
simple TTL/LVDS, ADC/DAC and motor encoder signals, here we focus on the
software communication (\figref{fwsw}).  PandABox's basic software interface,
\prog{PandABlocks-server}, communicates with the outside via TCP port 8888 for
control commands and TCP port 8889 for captured data; the former is taken care
by the pure-Python \verb|pandablocksclient.py|, and the latter by the EPICS
module \prog{ADPandABlocks} based on \prog{areaDetector} \cite{rivers2018}.
Correspondingly, in a \prog{Bluesky}-based environment, \prog{ADPandABlocks}
can be controlled with \prog{ophyd}'s built-in modules for \prog{areaDetector},
while \verb|pandablocksclient.py| needs to be encapsulated into
\prog{ophyd} with a new module (\figref{pbccomp}).

\begin{figure}[htbp]\centering
\includegraphics[width = 0.6\textwidth]{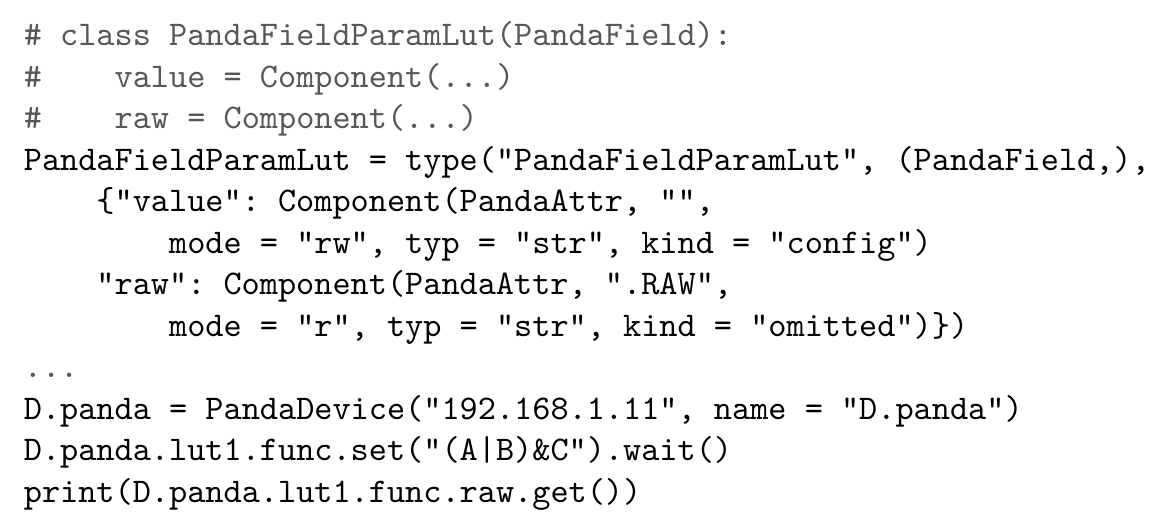}
\caption{%
	Example \prog{ophyd} class construction for a PandABox field,
	and example usage of our \prog{ophyd} module involving this field
}\label{fig:ophyd}
\end{figure}

A first issue we encounter when attempting to encapsulate PandABox's interface
into \prog{ophyd} is the variability of the list of configurable parameters,
determined by the firmware variant (and even version) on a particular ``PandA
box''. Fortunately, with \verb|pandablocksclient.py| (and indirectly the
controlling protocol of \prog{PandABlocks-server}), we can obtain the list
of PandABox's blocks, fields and attributes; furthermore, they are organised
under a succinct ``type system'' \cite{abbott2016}.  With this and Python's
\verb|type()| primitive, we can dynamically construct the \prog{ophyd} object
for a particular PandABox (\figref{ophyd}).  Here we note that \verb|type()| is
also employed in some dynamic constructs in \prog{ophyd} itself, and that what
we do is in essence similar to the handling of lists of camera features by
EPICS's \prog{ADGenICam} module \cite{rivers2019}; the latter module is the
basis of EPICS's unified support for GigE Vision and USB3 Vision cameras.

\begin{figure}[htbp]\centering
\includegraphics[width = 0.67\textwidth]{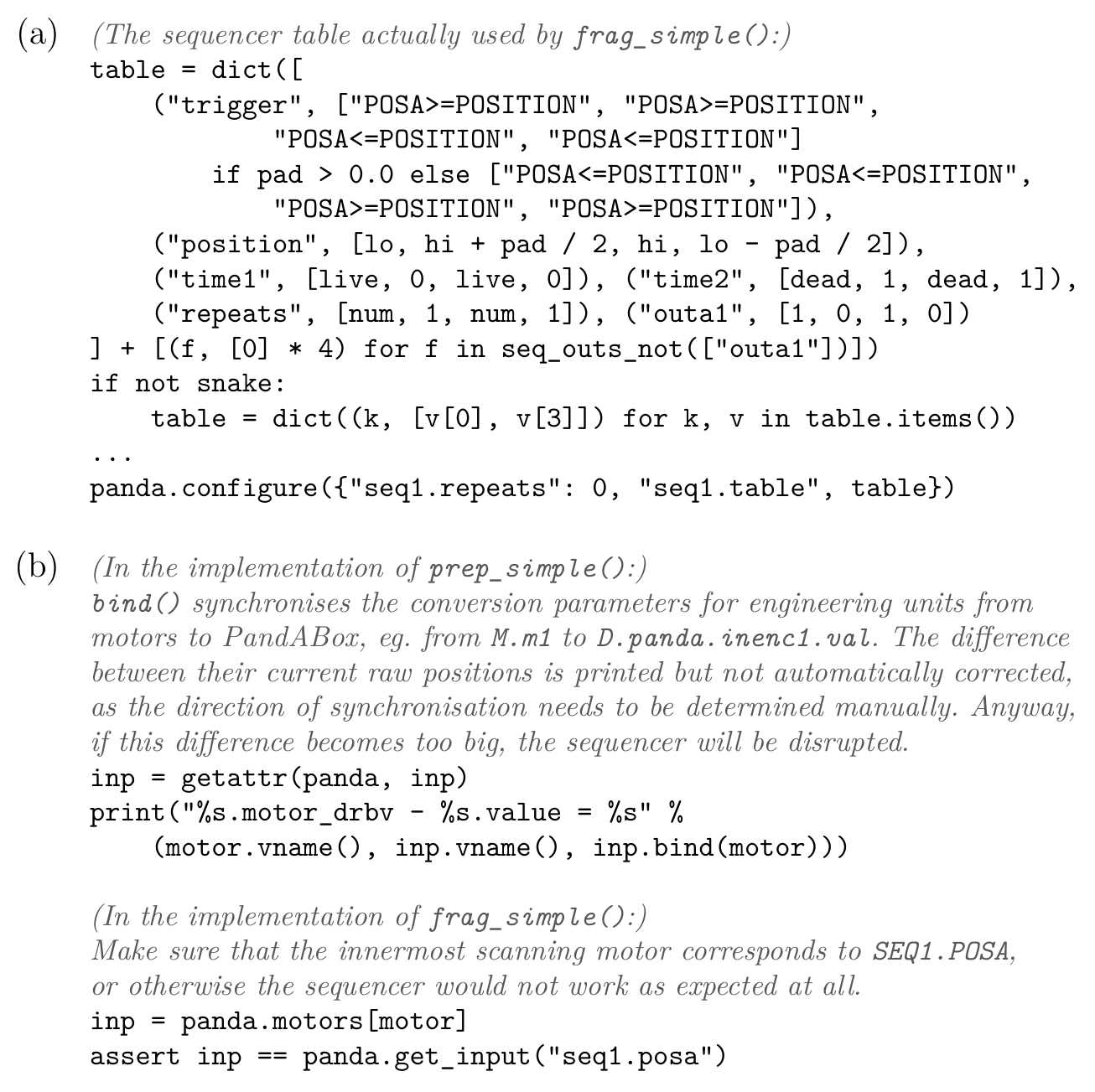}
\caption{%
	Useful extra facilities provided by our \prog{ophyd} module for PandABox:
	(a) table formatting (\cf\ \cite{dls2015b} for the sequencer
	semantics) and (b) motor binding (\cf\ \figref{seqplan}
	for the importance of the raw-position difference)%
}\label{fig:oextra}
\end{figure}

Our \prog{ophyd} module for PandABox is less than 380 lines of code, and has
developed to the point that we cannot find code we are obviously able to remove;
that is to say, it has approximated the \emph{complexity lower-bound} for its
purpose, and to us this means it has approached perfection as an EPG module
\cite{bentley1999}.  Actually there are two parts in this module which are not
strictly necessary for barebone hardware abstraction of PandABox, but they help
to simplify upper-level code so much that their existence is well justified
(\figref{oextra}).  One of them is table formatting facilities, thanks to which
we can specify sequencer tables and other tables succinctly and clearly; the
other is the binding between motor encoders and their ports on PandABox, which
will be discussed in \secref{epg}.  We also note that the motor-binding facility
in our module reimplements the similar facility in \prog{ADPandABlocks} in
a codebase less than $1/10$ the size of the latter's corresponding part.

\section{An EPG for $n$-dimensional constant-speed scans}\label{sec:epg}

\begin{figure}[htbp]\centering
\includegraphics[width = 0.8\textwidth]{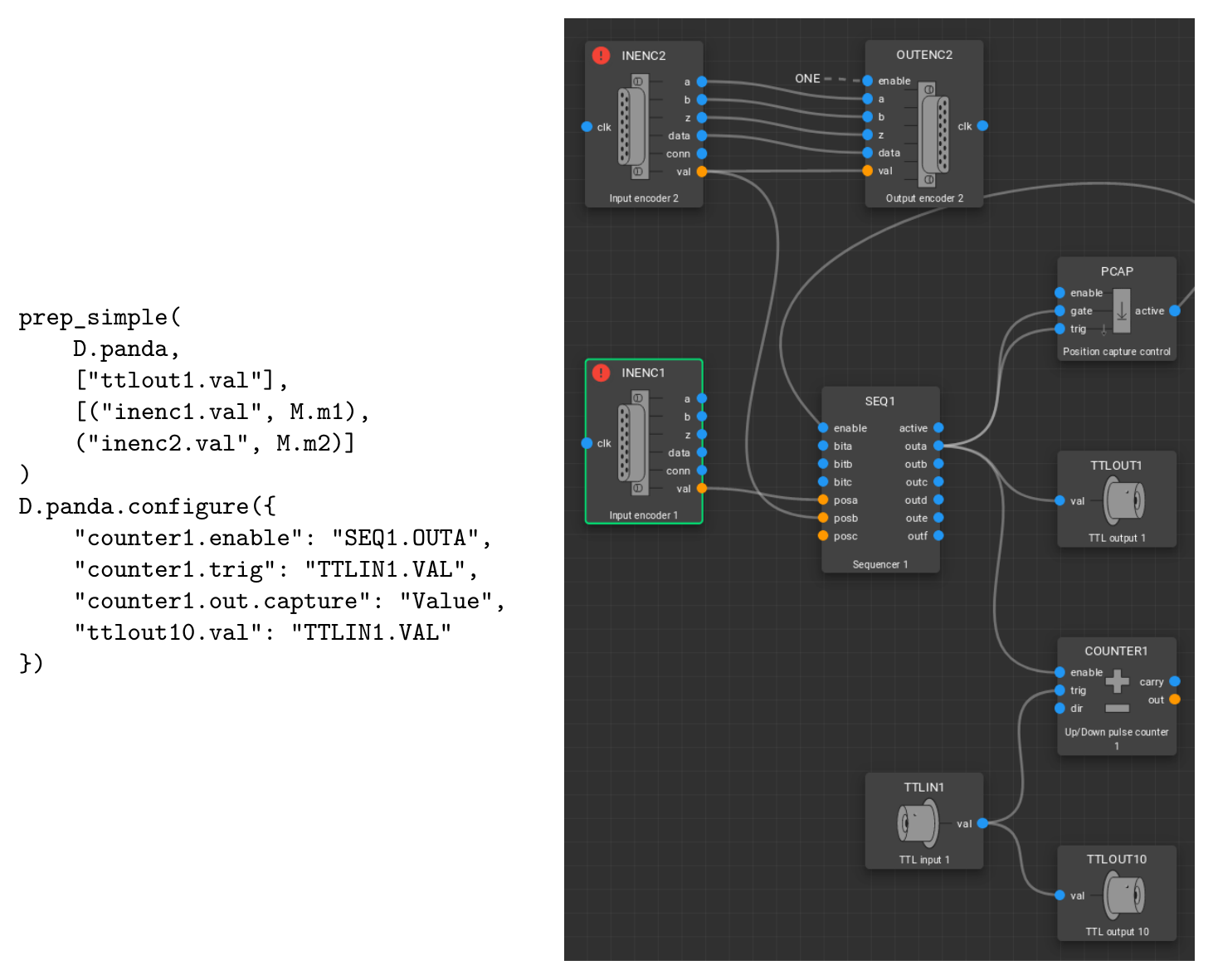}
\caption{%
	Configuration of PandA blocks actually used
	in XRF mapping at the 4W1B beamline of BSRF%
}\label{fig:blocks}
\end{figure}

From a programmer's perspective, a fly scan is mainly defined by the
trajectories of motors involved, as well as the corresponding rule for detector
triggering and data capture.  The former is implemented with profile moving of
motors in many scans, but can also be implemented by simple moves in the case
of constant-speed mapping.  The latter is, in PandABox-based scans, defined by
the configuration of PandA blocks, most importantly the wiring between blocks
as well as the sequencer tables used.  In this section we first discuss
the latter, and then discuss the former.  The first elements we care about
in a block configuration, apart from the capture control and sequencer blocks,
are the inputs and the outputs -- position information (whether as triggering
criteria or just as fields to be captured) and trigger outputs.  Following
this line of reasoning, the function \verb|prep_simple()| is provided by our
fly-scan backend EPG to simplify the typical configuration of PandA blocks
(\figref{blocks}).

With \verb|prep_simple()|, basic configuration is first applied to the blocks
\verb|PCAP| and \verb|SEQ1|, so that their activation/gating states are
associated in the typically expected way.  \verb|SEQ1|'s \verb|OUTA| and
inputs are then wired to, respectively, receivers of the trigger signals and
the specified encoder input blocks; the inputs blocks are additionally bound to
actual motor objects (\eg\ \verb|M.m1|, \verb|M.m2|) to facilitate some basic
sanity checks (\figref{oextra}).  Later in the function \verb|frag_simple()|,
\verb|SEQ1.POSA| is assumed to be wired to the innermost motor, which should
always corresponds to the ``flying'' axis.  Our EPG also takes care of the
difference between PandABox's encoder-card types automatically: the inputs and
outputs of ``monitor cards'' (\eg\ \verb/(IN|OUT)ENC1/ in \figref{blocks}) are
hard-wired together, while those of ``control cards'' (\eg\ \verb/(IN|OUT)ENC2/
in \figref{blocks}) are not; \verb|prep_simple()| automatically mimics the
behaviour of monitor cards on control cards by soft-wiring (which can of course
be overridden if desirable).  For XRF mapping at the 4W1B beamline of BSRF,
the $I_0$ value, V/F-converted to TTL signals, also needs to be recorded,
so we do some additional configuration by direct \prog{ophyd} calls after
the \verb|prep_simple()| call; we also duplicate the TTL input to
another TTL output, which can be fed to an external Ortec 974A
counter for backward compatibility with the old step-scan mode.

\begin{figure}[htbp]\centering
\includegraphics[width = 0.75\textwidth]{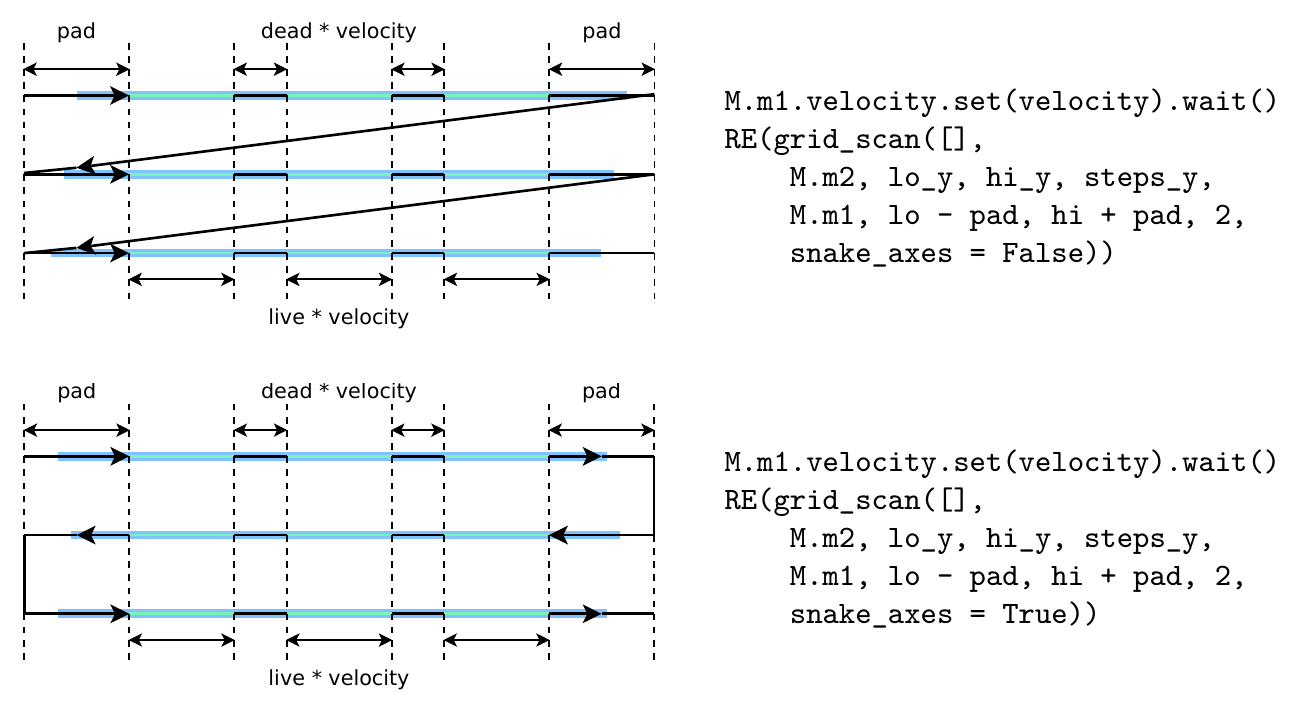}
\caption{%
	Sequencers (\cf\ implementation in \figref{oextra}(a)) and internally
	used \prog{Bluesky} calls (slightly modified) for raster and snake
	fly-scans: checkpoints for the sequencers are marked with arrow heads;
	the trajectories in an example condition with backlashes, excluding
	acceleration/deceleration phases, are shown in thick lines.%
}\label{fig:seqplan}
\end{figure}

As has been noted in the beginning of this section, the motion trajectories
in constant-speed mapping can be implemented by simple grid scans, which is
exactly what we do in our EPG (\figref{seqplan}), using \verb|grid_scan()| from
\prog{Bluesky}; of course, for the innermost motor, the number of points on each
line is hardcoded to 2, so that the motor moves continuously from the beginning
to the end.  To avoid troubles with acceleration/deceleration phases and
potential backlashes, \emph{pad intervals} (0.5 second for each interval by
default) are automatically added to both sides on the ``flying'' axes.  Assuming
an alternation between exposure (TTL high) of a fixed time and darkness (TTL
low) of another fixed time, the sequencer is pretty straightforward to design;
the potential issue with the speed (un)stability of motors will be briefly
discussed in \secref{bextra}.  We also note that in order to prevent premature
starting of sequence for the next line after end of the current line, a
deadband mechanism is implemented: the next-line sequence is not enabled
until the innermost motor passes a checkpoint, which is the centre of the
pad interval.  With all these elements available, we are able to provide a
command-line interface for constant-speed mapping fly scans (\figref{fragplan};
\verb|fly_demo()| is the \prog{Bluesky} ``plan'' used for XRF at 4W1B of BSRF)
similar to the corresponding step-scan interface provided by \prog{Bluesky}.
Only two additional parameters are needed: the length of each exposure-darkness
period in seconds, and the duty ratio of exposure in each of these periods.

\section{Backend extension and further considerations}\label{sec:bextra}

\begin{figure}[htbp]\centering
\includegraphics[width = 0.7\textwidth]{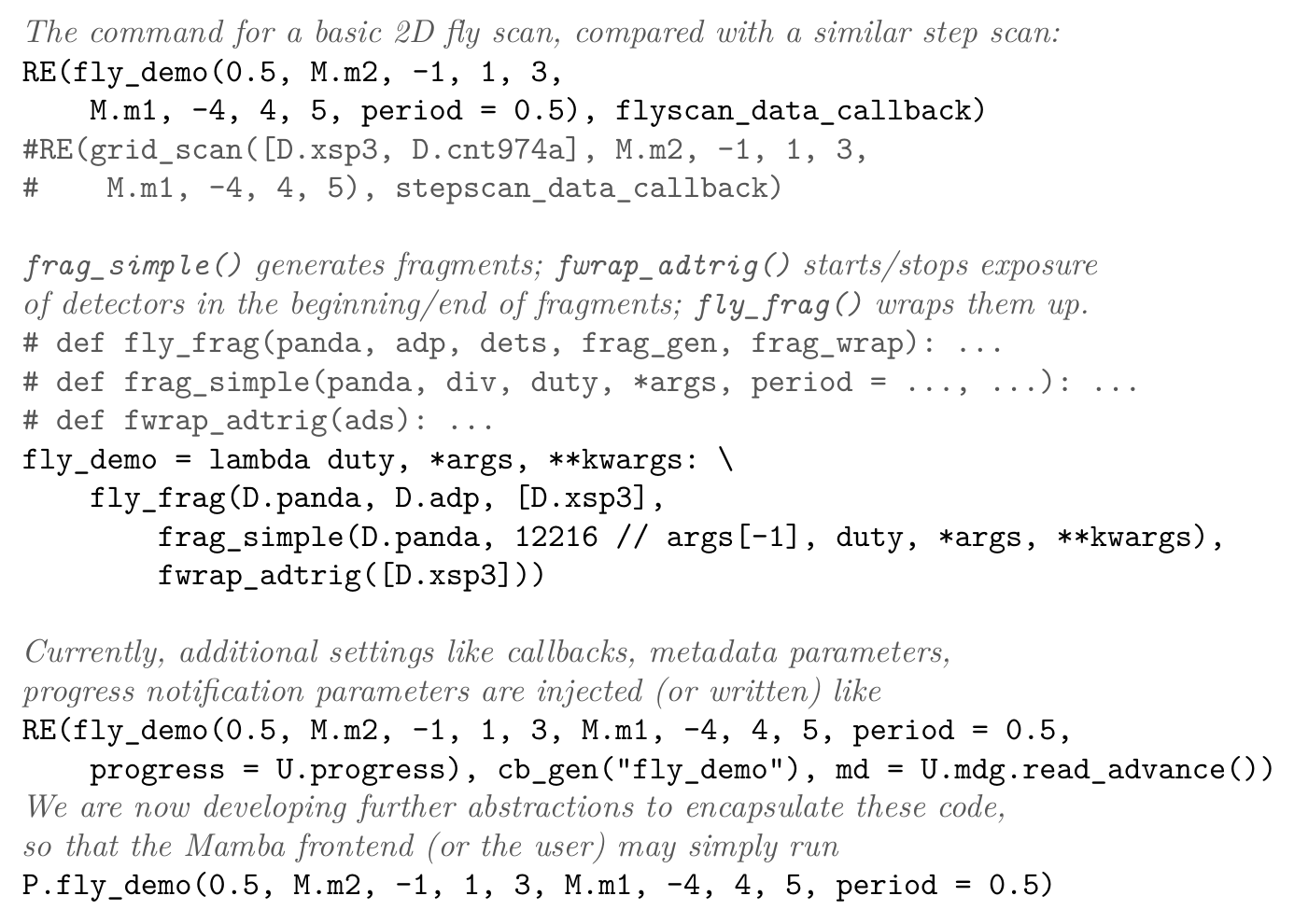}
\caption{%
	Example \prog{Bluesky} command used by command-line users
	and \prog{Mamba} frontends for constant-speed scans,
	and an outline of its internal structure%
}\label{fig:fragplan}
\end{figure}

One issue that may be encountered in fly scans with many scan points is
hardware limits on the number of frames in one exposure series: \eg\ the
Xspress3 readout system for silicon drift detectors (SDDs), used for XRF
experiments at the 4W1B beamline of BSRF, often has a limit of 12216
frames; PandABox itself also has a limit of 4096 entries in a sequencer
table, which may become a problem with scans involving complex sequencers.
\prog{pymalcoms}'s official way to deal with the latter is to use a
double-buffer design \cite{dls2015c}, where the sequencers \verb|SEQ1| and
\verb|SEQ2| in PandABox are used by turn, and new table entries are filled
into a sequencer after the latter has been exhausted.  At HEPS, we have
explored what we call \emph{scan fragmentation} (\figref{fragplan}) as a
generic solution: big scans are programmatically divided into fragments
that do not exceed the hardware limits involved; the EPG takes care of
the stopping/restarting of exposure between fragments and the stitching of
data streams from them, so that the user just sees an uninterrupted scan.

In our backend EPG, the upper-level code is less than 190 lines, and is quite
extensible.  \emph{Eg.}\ as has been suggested in \secref{epg}, implementing
irregular fly scans or the double-buffer design mentioned above (should
it be necessary for some big ultrafast scans) is essentially a matter of
implementing new block configurations (most importantly wiring and sequencer
tables) and new motion trajectories.  Although the implementational complexity
will surely be non-trivial, architecturally it will be totally compatible with
our current framework, and hopefully we will still be able to approximate the
complexity lower-bounds with this framework.  An easier extension is real
position-based triggering (perhaps with PandABox's \verb|PCOMP| blocks),
instead of triggering with even temporal spacing; however, with decent motors
the speeds are usually very stable: \eg\ for the sample-stage motors with
Kohzu ARIES controllers at the 4W1B beamline of BSRF, we find the nonlinearlity
of the position curves to be roughly 0.1\%, which is virtually negligible.
On the other hand, for less stable motors and even deliberately irregular
scans (\eg\ those used in ptychography), the problem with uneven spacing
of scan points can also be eliminated by plotting the results
as Voronoi diagrams \cite{weisstein2022}.

Architecturally, our EPG framework is also evolving.  First of all, with the
SWMR (single writer, multiple reader) extension of HDF5, we are already able
to do online processing of data acquired from the scans nearly independent
of \prog{Bluesky}'s data-processing facilities (\cf\ \secref{frontend}).
However, the messages generated by \prog{Bluesky}'s \verb|RunEngine|, \eg\ %
\verb|start|, \verb|stop| and \verb|event|, can still act as progress markers
and metadata carriers, which are of great help in sanity checks and applications
like ETA estimation; this is why \prog{Bluesky}'s callbacks are still used
by \prog{Mamba} for fly scans.  To deal with further real (and reasonable)
requirements for HEPS and in pilot tests at BSRF, we have developed features
like advanced metadata management, ETA estimation \etc\ in \prog{Mamba}.
As a result, we find the commands injected by the \prog{Mamba} frontend into
the backend more and more bloated (\figref{fragplan}); we are now extending
our framework, so that we may replace boilerplate code with simple calls to
wrappers like \verb|P.fly_demo()| (\verb|P| for ``plans'', like the \verb|U|
for ``utilities'' in the same figure, also introduced by \prog{Mamba}),
which is what we believe an EPG's interface should look like.

\section{A \prog{Mamba} frontend for XRF mapping}\label{sec:frontend}

\begin{figure}[htbp]\centering
\includegraphics[width = 0.6\textwidth]{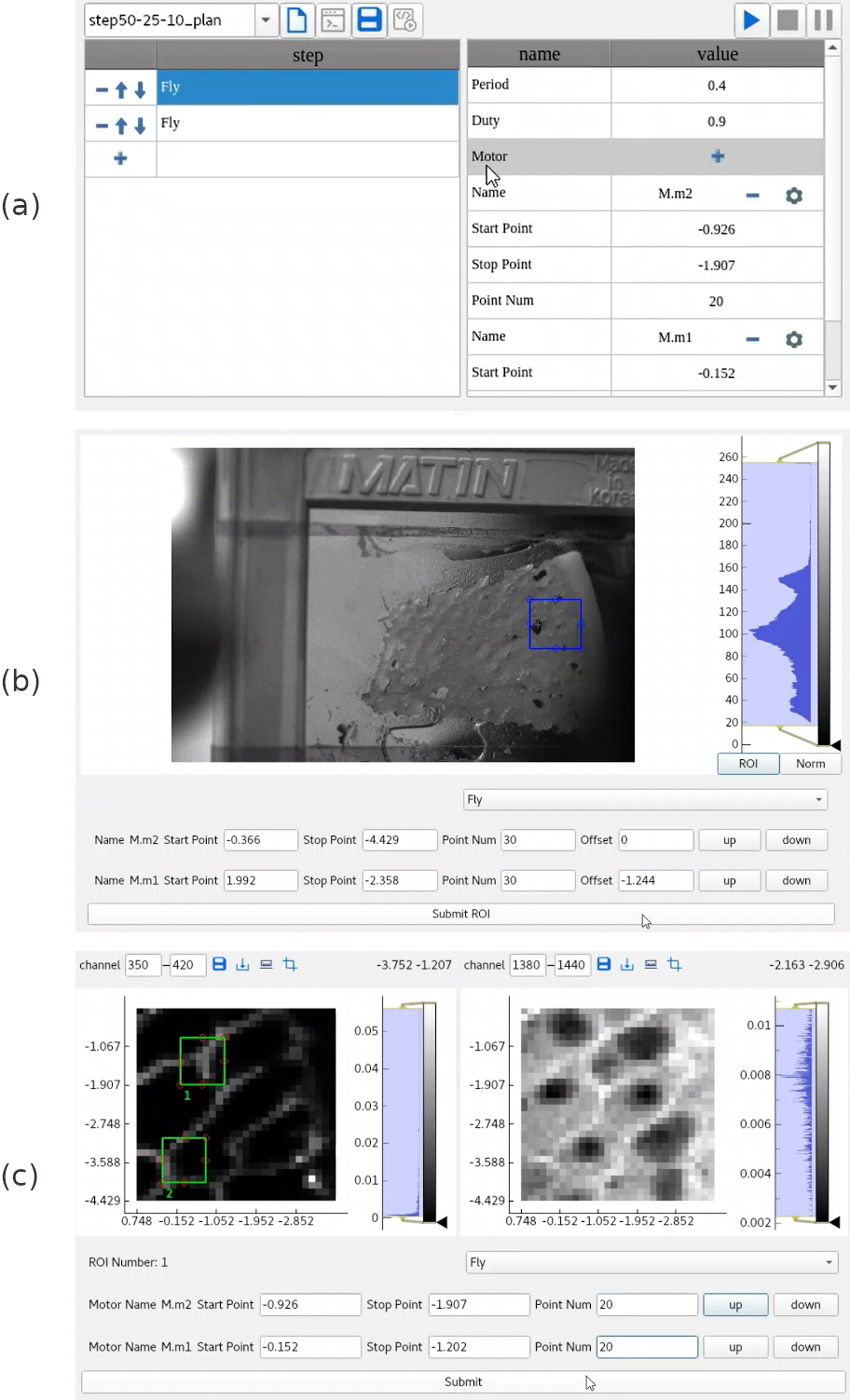}
\caption{%
	\prog{Mamba}'s (a) scan sequence, (b) ROI
	finder and (c) 2D element visualisation modules;
	the last is shown with the ROI selection tool.%
}
\label{fig:seqroi}
\end{figure}

Based on our backend EPG, a \prog{Mamba} GUI has been developed to configure
experimental procedures and monitor data acquisition in XRF mapping.  To
accommodate for various step-scan and fly-scan requirements, an intuitive
scan sequence module (\figref{seqroi}(a)) is provided, where acquisition
sequences can be specified for the batch scanning of multiple ROIs.
To further facilitate the selection of scan regions, an ROI finder plugin
(\figref{seqroi}(b)) is developed to select scan regions from a monitoring
camera.  Otsu's algorithm \cite{otsu1979} is used to track the movement of a
reference object in acquired images, and to calculate the conversion parameters
between the pixel coordinates and the motors' engineering coordinates; then scan
regions can be selected from the ROI finder and automatically passed to the scan
sequence module.  The scan sequence module also integrates with the 2D element
visualisation module: ROIs can be extracted from the acquired element images
in the responsive visualisation tool (\figref{seqroi}(c)), and the corresponding
coordinates are also passed to the scan sequence module for future scans,
furtherly enhancing the user-friendliness of fly scans with \emph{Mamba}.

\begin{figure}[htbp]\centering
\includegraphics[width = 0.5\textwidth]{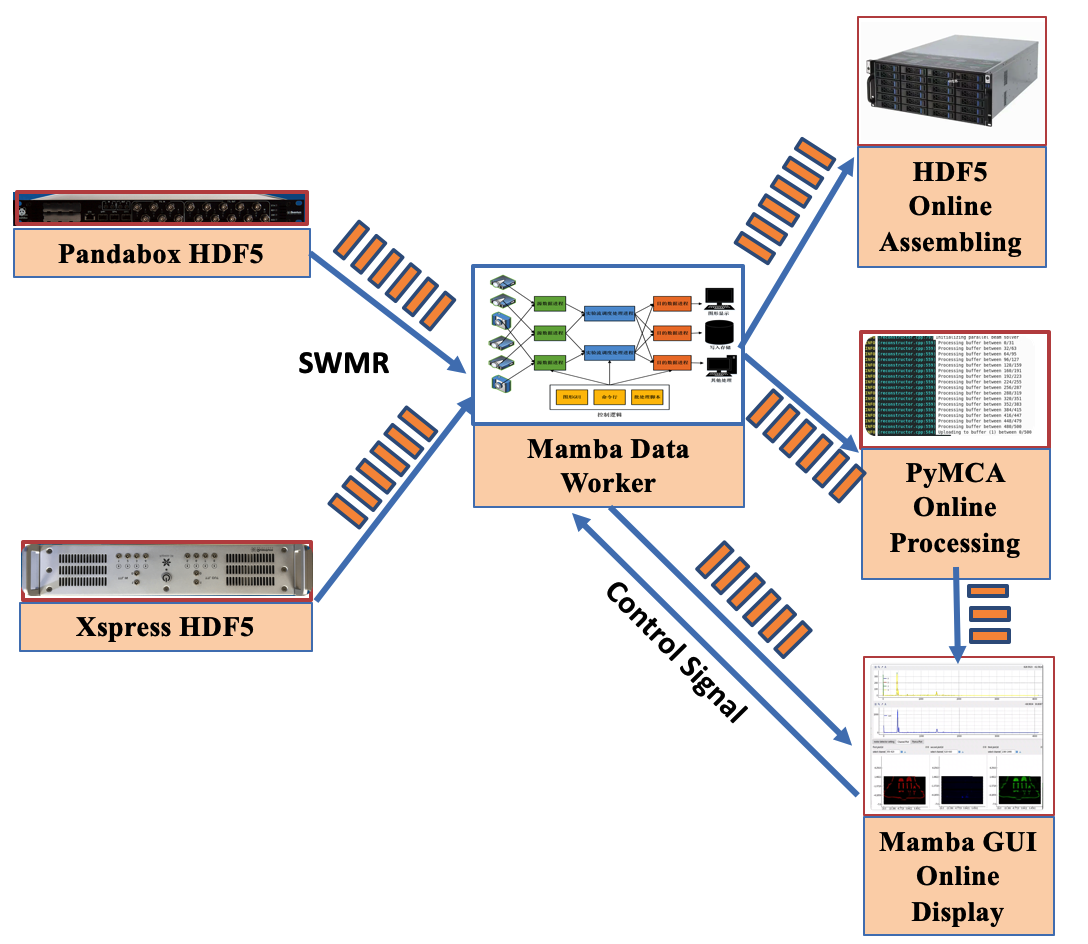}
\caption{\prog{Mamba}'s dataflow graph for XRF mapping at 4W1B of BSRF}
\label{fig:dataflow}
\end{figure}

In XRF mapping, the online visualisation of element distributions can help
to extract ROIs from large samples and to further adjust acquisition parameters,
which are essential in improving the efficiency of XRF mapping experiments.
The online data analysis of acquired spectra is performed by the dataflow
component \prog{Mamba Data Worker}; the \prog{PyMca} toolkit is integrated
into the dataflow graph to obtain element distributions.  In fly scans at
the 4W1B beamline of BSRF, PandABox and Xspress3 write into two separate
HDF5 files, both with SWMR enabled; \prog{Mamba Data Worker} monitors the
HDF5 files and assembles freshly generated data into a data block when
a threshold is reached.  Data blocks are multiplexed to three receivers
(\figref{dataflow}): raw-data storage (in the form of HDF5/NeXus files with
acquisition metadata included), \prog{PyMca} processing and the \prog{Mamba}
GUI.  The processing of acquired spectra can be tuned online from the GUI,
offering users more flexibility; the \prog{PyMca} workflow is also extended
with a linear fitting mode to enable fast feedback after this kind of tuning.

\section{Conclusion}

By reusing the module \verb|pandablocksclient.py| from \prog{pymalcolm}, the
interface of PandABox has been encapsulated in an \prog{ophyd} module; by fully
employing PandABox's ability to output its list of configurable parameters and
the ``type system'' behind these parameters, we are able to cleanly handle the
variability of PandABox's interface.  Our \prog{ophyd} module for PandABox has
approximated the complexity lower-bound for its purpose, and also provides
helpful facilities for PandA table formatting and motor binding.  Based on this
module, we provided a utility function to simplify the typical configuration of
PandA blocks, as well as utility functions to generate the motion trajectories
and sequencers for $n$-dimensional constant-speed mapping.  Our generators also
support a flexible scan fragmentation mechanism to deal with limits on number
of exposure frames or sequencer table entries imposed by hardware.  Our
\prog{Bluesky}-based fly-scan framework can be cleanly extended to more complex
(irregular, ultrafast \etc) scans, and we are also continuously improving
our framework to handle more complex requirements at the same time,
aside from fly scans.  Based on our framework, a user-friendly
\prog{Mamba} frontend is developed for XRF mapping experiments,
which provides fully online visual feedback.

\section*{Statements and Declarations}

\paragraph{Acknowledgements:}
All authors of this paper gratefully acknowledge the 4W1B and 3W1 beamlines
of BSRF for providing software testing beamtime.  The authors acknowledge
the National High Energy Physics Datacenter and High Energy Physics
Datacenter of Chinese Academy of Sciences for providing computing and
storage resources.

\paragraph{Funding:}
This work was supported by the National Science Foundation for
Young Scientists of China (Grants No.\ 12005253 and No.\ 12205328),
the Strategic Priority Research Program of Chinese Academy of Sciences
(XDB37000000) and the Technological Innovation Program of Institute
of High Energy Physics of Chinese Academy of Sciences (E25455U210).

\paragraph{Data Availability:}
A fully open-source edition of \emph{Mamba} is available at
\url{https://github.com/CasperVector/mamba-ose};
it depends on currently HEPS-specific patches for
\emph{Bluesky} components, and these patches are available at
\url{https://github.com/CasperVector/ihep-pkg-ose/tree/master/misc/pybuild}.

\bibliography{art6}
\end{document}